\documentclass[twocolumn, prX, showkeys]{revtex4}
\usepackage{amsmath,amssymb,multirow,amsfonts,amsthm}
\usepackage[pdftex]{graphicx}
\usepackage[tight]{subfigure}
\usepackage[all]{xypic}

\begin{document}
	
		\title{Rare Event Simulation}
		
		\author{James L. Beck}
		\affiliation{California Institute of Technology, Pasadena, CA 91125}
		\email{jimbeck@caltech.edu}
		\author{Konstantin M. Zuev}
		\affiliation{Northeastern University, Boston, MA 02115}
		\email{k.zuev@neu.edu}
		
		\begin{abstract}
			Rare events are events that are expected to occur infrequently, or more technically, those that have low probabilities (say, order of $10^{-3}$ or less) of occurring according to a probability model. In the context of uncertainty quantification, the rare events often correspond to failure of systems designed for high reliability, meaning that the system performance fails to meet some design or operation specifications. As reviewed in this section, computation of such rare-event probabilities is challenging. Analytical solutions are usually not available for non-trivial problems and standard Monte Carlo simulation is computationally inefficient. Therefore, much research effort has focused on developing advanced stochastic simulation methods that are more efficient. In this section, we address the problem of estimating rare-event probabilities by Monte Carlo simulation, Importance Sampling and Subset Simulation for highly reliable dynamic systems.
		\end{abstract}
		
		
		\maketitle

\section{Introduction}
We focus on rare-event simulation for addressing reliability problems corresponding to dynamic systems. To compute the rare event (failure) probability for a dynamic system, both input (excitation) and modeling uncertainties should be quantified and propagated. Therefore, a probability model must be chosen to describe the uncertainty in the future input for the system and then a chosen deterministic or stochastic system model is used, preferably in conjunction with a probability model describing the associated modeling uncertainties, to propagate these uncertainties. These input and system models define a probabilistic description of the system output (response). For example, the problem of interest might be to compute the small failure probability for a highly reliable dynamic system such as a bridge or building under uncertain future earthquake excitation, or for an aircraft under uncertain excitation by turbulence, using a finite-element structural model to approximate the dynamics of the system. This model will usually be subject to both parametric uncertainty (what values of the model parameters best represent the behavior of the system?) and non-parametric modeling uncertainty (what are the effects of the aspects of the system behavior not captured by the dynamic model?). The treatment of input uncertainty has a long history in dynamic reliability theory and random vibrations, now more commonly called stochastic dynamics, but the treatment of modeling uncertainty is more recent.

Usually the dynamic model of the system is represented by a time-dependent BVP (boundary-value problem) involving PDEs (partial differential equations) or by a set of coupled ODEs (ordinary differential equations). Typically the failure event is defined as any one of a set of performance quantities of interest exceeding its specified threshold over some time interval.  This is the so-called \textit{first-passage problem}. This challenging problem is characterized by a lack of analytical solutions, even for the simplest case of a single-degree-of-freedom linear oscillator subject to excitation that is modeled as a Gaussian process. Approximate analytical methods exist that are usually limited in scope and their accuracy is difficult to assess in a given application \cite{Rackwitz,Taflanidis_Beck}. Semi-analytical methods from structural reliability theory such as FORM and SORM (first- and second-order reliability methods) \cite{Ditlevsen,Rackwitz} cannot be applied directly to the first-passage problem and are inapplicable, anyway, because of the high-dimensional nature of the discrete-time input history \cite{KatZuev,Valdebenito}. Standard Monte Carlo simulation has general applicability but it is computationally very inefficient because of the low failure probabilities. As a consequence, advanced stochastic simulation schemes are needed.

\subsection{Mathematical formulation of problem}

We assume that initially there is a continuous-time deterministic model of the real dynamic system that consists of a state-space model with a finite-dimensional state $X(t)\in\mathbb{R}^n$  at time $t$ and this is converted to a discrete-time state-space model using a numerical time-stepping method to give:
\begin{equation}\label{model}
\begin{split}
&X(t+1)=f(X(t),U(t),t),\\
&X(t)\in\mathbb{R}^n, \hspace{3mm} U(t)\in\mathbb{R}^m, \hspace{3mm} t=0,\ldots,T,
\end{split}
\end{equation}
where $U(t)\in\mathbb{R}^m$ is the input at discrete time $t$.

If the original model consists of a BVP with PDEs describing a response $u(x,t)$ where $x\in\mathbb{R}^d$, then we assume that a finite set of basis functions $\{\phi_1(x),\ldots,\phi_n(x)\}$  is chosen (e.g. global bases such as Fourier and Hermite polynomials or localized ones such as finite-element interpolation functions) so that the solution is well approximated by:
\begin{equation}
u(x,t)\approx\sum_{i=1}^nX_i(t)\phi_i(x).
\end{equation}
Then a numerical method is applied to the BVP PDEs to establish time-dependent equations for the vector of coefficients $X(t)=[X_1(t),\ldots, X_n(t)]$ so that the standard state-space equation in (\ref{model}) still applies. For example, for a finite-element model of a structural system, $\{\phi_1(x),\ldots,\phi_n(x)\}$   would be local interpolation functions over the elements. Then, expressing the BVP in weak form, a weighted residual or Galerkin method could be applied to give a state-space equation for the vector of coefficients $X(t)$ \cite{Johnson}.

Suppose that a positive scalar performance function $g(X(t))$ is a quantity of interest and that the rare event $\mathcal{E}$ of concern is that $g(X(t))$ exceeds a threshold $b$ over some discrete-time interval $t=0,\ldots,T$: 
\begin{equation}\label{F}
\begin{split}
&\mathcal{E}=\left\{U=(U(0),\ldots,U(T)):\hspace{-1mm}\max\limits_{t=0,\ldots,T}g(X(t))> b\right\},\\
&\mathcal{E}\subset\mathbb{R}^{m\times(T+1)},
\end{split}
\end{equation}
where $X(t)$ satisfies (\ref{model}). The performance function $g(X(t))$ may involve exceedance of multiple performance quantities of interest $\{g_k(X(t)): k=1,\ldots,K\}$ above their corresponding thresholds $\{a_k\}$. This can be accomplished by aggregating them using the \textit{max} and \textit{min} operators in an appropriate combination on the set of $g_k$'s; for example, for a \textit{pure series} failure criterion, where the threshold exceedance of \textit{any} $a_k$ represents failure, one takes the aggregate performance failure criterion as $g(X(t)) = \max \{g_k(X(t))/a_k: k = 1,\ldots,K\}>1$, while for a \textit{pure parallel} failure criterion, where \textit{all} of the $g_k$ must exceed their thresholds before failure is considered to have occurred, one takes the aggregate performance failure criterion as $g(X(t)) = \min \{g_k(X(t))/a_k: k=1,\ldots,K\} >1$.

If the uncertainty in the input time history vector $U=[U(0),\ldots, U(T)]\in\mathbb{R}^D$  $(D=m\times(T+1))$ is quantified by a probability distribution for $U$ that has a PDF (probability density function) $p(u)$ with respect to Lebesgue integration over $\mathbb{R}^D$, then the rare-event probability is given by: 
\begin{equation}\label{pF}
p_\mathcal{E}=\mathbb{P}(U\in \mathcal{E})=\int_\mathcal{E} p(u)du.
\end{equation}
The PDF $p(u)$ is assumed to be readily sampled. Although \textit{direct sampling} from a
high-dimensional PDF is not possible in most cases, multi-dimensional Gaussians are an exception because the Gaussian vector can be readily transformed so that the components are independent and the PDF is a product of one-dimensional Gaussian PDFs. In many applications, the
discrete-time stochastic input history is modeled by running discrete-time Gaussian white noise through a digital filter to shape its spectrum in the frequency domain, and then multiplying the filtered sequence by an envelope function to shape it in the time domain, if it is non-stationary.

The model in (\ref{model}) may also depend on uncertain parameters $\theta\in\Theta\subset\mathbb{R}^p$ which includes the initial values $X(0)$ if they are uncertain. Then a prior PDF $p(\theta)$ may be chosen to quantify the uncertainty in the value of vector $\theta$. Some of the parameters may characterize the PDF for input $U$ which can then be denoted $p(u|\theta)$. It is convenient to re-define vector $U$ to also include $\theta$, then the new PDF $p(u)$ is $p(u|\theta) p(\theta)$ in terms of the previous PDFs. We assume that model parameter uncertainty is incorporated in this way, so the basic equations remain the same as (\ref{model}), (\ref{F}), and (\ref{pF}). When model uncertainty is incorporated, the calculated $p_\mathcal{E}$ has been referred to as the robust rare-event probability \cite{PBK01,Be10}, meaning robust to model uncertainty, as in robust control theory.

\section{Standard Monte Carlo Simulation}
The standard \textit{Monte Carlo Simulation} method (MCS) is one of the most robust and straightforward ways to simulate rare events and estimate their probabilities. The method was originally developed in \cite{MonteCarlo} for solving problems in mathematical physics. Since then MCS has been used in many  applications in physics, statistics, computer science, and engineering, and currently it lays at the heart of all random sampling-based techniques \cite{Liu,Robert}. 

The basic idea behind MCS is to observe that the probability in (\ref{pF}) can be written as an expectation:
\begin{equation}\label{pF=E[I]}
p_\mathcal{E}=\int_{\mathbb{R}^{D}}{I_\mathcal{E}(u)p(u)du}=\mathbb{E}_p[I_\mathcal{E}], 
\end{equation}
where $I_\mathcal{E}$ is the indicator function of $\mathcal{E}$, that is $I_\mathcal{E}(u)=1$ if $u\in \mathcal{E}$ and $I_\mathcal{E}(u)=0$ otherwise, and $D=m\times(T+1)$ is the dimension of the integral. Recall that the strong law of large numbers \cite{Ross} states that if
$U_1,\ldots,U_N$ are independent and identically
distributed (i.i.d.) samples of vector $U$ drawn from the distribution $p(u)$, then for any function $h(u)$ with finite mean $\mathbb{E}_{p}[h(u)]$, the sample average
$\frac{1}{N}\sum_{i=1}^Nh(U_i)$ converges to the true value
$\mathbb{E}_{p}[h(u)]$ as $N\rightarrow\infty$ almost surely (i.e. with probability 1). Therefore, setting $h(u)=I_{\mathcal{E}}(u)$, the probability in (\ref{pF=E[I]}) can be estimated as follows:
\begin{equation}\label{MCSestimate}
p_\mathcal{E}\approx {p}_\mathcal{E}^{MCS}=\frac{1}{N}\sum_{i=1}^NI_\mathcal{E}(U_i).
\end{equation}
It is straightforward to show that ${p}_\mathcal{E}^{MCS}$ is an \textit{unbiased} estimator of $p_\mathcal{E}$ with mean and variance:
\begin{equation}\label{mean_var}
\begin{split}
\mathbb{E}_p[{p}_\mathcal{E}^{MCS}]&= \mathbb{E}_p\left[\frac{1}{N}\sum_{i=1}^NI_\mathcal{E}(U_i)\right]\\
&=
\frac{1}{N}\sum_{i=1}^N\mathbb{E}_p[I_\mathcal{E}]=p_\mathcal{E},\\
\mathrm{Var}_p[{p}_\mathcal{E}^{MCS}]&=\mathrm{Var}_p\left[\frac{1}{N}\sum_{i=1}^NI_\mathcal{E}(U_i)\right]\\
&=\frac{1}{N^2}\sum_{i=1}^N\mathrm{Var}_p[I_\mathcal{E}]=\frac{p_\mathcal{E}(1-p_\mathcal{E})}{N}.
\end{split}
\end{equation}
Furthermore, by the Central Limit Theorem \cite{Ross}, as $N\rightarrow\infty$, ${p}_\mathcal{E}^{MCS}$ is distributed asymptotically as Gaussian with this mean and variance.

\textit{Frequentist interpretation of MCS:}
 The frequentist interpretation of MCS focuses on the forward problem, arguing that if $N$ is large so that the variance of ${p}_\mathcal{E}^{MCS}$ is relatively small, then the value $\widehat{p}_\mathcal{E}^{MCS}$ based on (\ref{MCSestimate}) for a specific set of $N$ samples $\{\widehat{U}_1,\ldots,\widehat{U}_N\}$ drawn from $p(u)$ should be close to the mean $p_\mathcal{E}$ of ${p}_\mathcal{E}^{MCS}$. The sample mean estimate $\widehat{p}_\mathcal{E}^{MCS}$ is very intuitive and, in fact, simply reflects the frequentist definition of probability: $\widehat{p}_\mathcal{E}^{MCS}$ is the ratio between the number of trials where the event $\mathcal{E}$ occurred, $\widehat{N}_\mathcal{E}=\sum_{i=1}^NI_\mathcal{E}(\widehat{U}_i)$, and the total number of trials $N$.

\textit{Bayesian interpretation of MCS:} The same MCS estimate  $\widehat{p}_\mathcal{E}^{MCS}$  has a simple Bayesian interpretation (e.g. \cite{BSS}), which focuses on the inverse problem for the specific set of $N$ samples $\{\widehat{U}_1,\ldots,\widehat{U}_N\}$ drawn from $p(u)$. 
Following the Bayesian approach \cite{Jaynes_book}, the unknown probability $p_\mathcal{E}$ is considered as a stochastic variable whose value in $[0,1]$ is uncertain. The \textit{Principle
of Maximum Entropy} \cite{Jaynes} leads to the uniform prior distribution for $p_\mathcal{E}$, $p(p_\mathcal{E})=1$, $0\leq p_\mathcal{E}\leq1$, which implies that all values are taken as equally plausible a priori. Since samples $U_1,\ldots,U_N$ are i.i.d, the binary sequence $I_\mathcal{E}(U_1),\ldots,I_\mathcal{E}(U_N)$ is a sequence of Bernoulli trials, and so for the forward problem,  $N_\mathcal{E}$ is distributed according to the binomial distribution with parameters $N$ and $p_\mathcal{E}$, $N_\mathcal{E}\sim {\mathrm{Bin}}(N,p_\mathcal{E})$. Therefore, for the set of $N$ samples, the likelihood function is  $p(\widehat{N}_\mathcal{E}|p_\mathcal{E},N)={N \choose \widehat{N}_\mathcal{E}}p_{\mathcal{E}}^{\widehat{N}_\mathcal{E}}(1-p_\mathcal{E})^{N-\widehat{N}_\mathcal{E}}$. Using \textit{Bayes' Theorem}, the posterior distribution for $p_\mathcal{E}$, $p(p_\mathcal{E}|\widehat{N}_\mathcal{E},N)\propto p(p_\mathcal{E})p(\widehat{N}_\mathcal{E}|p_\mathcal{E},N)$,  is therefore the beta distribution $\mathrm{Beta}(\widehat{N}_\mathcal{E}+1,N-\widehat{N}_\mathcal{E}+1)$, i.e. 
\begin{equation}\label{posterior}
p(p_\mathcal{E}|\widehat{N}_\mathcal{E},N)=\frac{p_{\mathcal{E}}^{\widehat{N}_\mathcal{E}}(1-p_\mathcal{E})^{N-\widehat{N}_\mathcal{E}}}{B(\widehat{N}_\mathcal{E}+1,N-\widehat{N}_\mathcal{E}+1)},
\end{equation}
where the beta function $B$ is the normalizing constant that equals $(N+1)!/(\widehat{N}_\mathcal{E}!(N-\widehat{N}_\mathcal{E})!)$ here. The MCS estimate is the \textit{maximum a posteriori (MAP) estimate}, which is the mode of the posterior distribution (\ref{posterior}) and therefore the most probable value of $p_\mathcal{E}$ a posteriori:
\begin{equation}
\widehat{p}_\mathcal{E}^{MCS}=\frac{\widehat{N}_\mathcal{E}}{N}.
\end{equation}
Notice that the posterior PDF in (\ref{posterior}) gives a complete description of the uncertainty in the value of $p_\mathcal{E}$ based on the specific set of $N$ samples of $U$ drawn from $p(u)$. The posterior distribution in (\ref{posterior}) is in fact the original Bayes' result \cite{Bayes}, although  Bayes' Theorem was developed in full generality by Laplace \cite{Laplace1}.    

The standard MCS method for estimating the  probability in (\ref{pF})  is summarized in the following pseudo-code.  \\

\vspace{2mm} 
\hrule height 0.6pt \rule{0pt}{4mm}\centerline
{\textbf{Monte Carlo Simulation}}\vspace{.5mm}\rule{0pt}{4mm}
\hrule
\vspace{1mm}
\texttt{Input:}

\hspace{0.5cm}$\vartriangleright$ $N$, total number of samples.

\texttt{Algorithm:}

\hspace{0.5cm} Set $N_\mathcal{E}=0$, number of trials where the event $\mathcal{E}$ 

\hspace{0.5cm} occurred.

\hspace{0.5cm} \textbf{for} $i=1,\ldots,N$ \textbf{do}

\hspace{1.2cm}  Sample the input excitation 

\hspace{1.2cm} $U_i=(U_i(0),\ldots,U_i(T))\sim p(u)$.

\hspace{1.2cm} Compute the system trajectory 

\hspace{1.2cm} $X_i=(X_i(0),\ldots,X_i(T))$ 

\hspace{1.2cm} using the system model (\ref{model}) with $U(t)=U_i(t)$.

\hspace{1.2cm} \textbf{if}  $\max\limits_{t=0,\ldots,T}g(X_i(t))> b$

\hspace{1.6cm} $N_\mathcal{E}\leftarrow N_\mathcal{E}+1$

\hspace{1.2cm} \textbf{end if}

\hspace{0.5cm} \textbf{end for}

 \texttt{Output:}
 
 \hspace{0.5cm}$\blacktriangleright$  $\widehat{p}_\mathcal{E}^{MCS}=\frac{N_\mathcal{E}}{N}$, MCS estimate of $p_\mathcal{E}$
 
 \hspace{0.5cm}$\blacktriangleright$ $p(p_\mathcal{E}|N_\mathcal{E},N)=\frac{p_{\mathcal{E}}^{N_\mathcal{E}}(1-p_\mathcal{E})^{N-N_\mathcal{E}}}{B(N_\mathcal{E}+1,N-N_\mathcal{E}+1)}$, posterior PDF 
 
 \hspace{0.5cm} of $p_\mathcal{E}$
 
  \vspace{1mm}
  \hrule
  \vspace{5mm}
  
\textit{Assessment of accuracy of MCS estimate:}  For the frequentist interpretation, the \textit{coefficient of variation} (c.o.v.) for the estimator ${p}_\mathcal{E}^{MCS}$ given by (\ref{MCSestimate}), conditional on $p_\mathcal{E}$ and $N$, is given by (\ref{mean_var}):
\begin{equation}\label{cov}
\delta({p}_\mathcal{E}^{MCS}|p_\mathcal{E},N)=\frac{\sqrt{\mathrm{Var}_p[{p}_\mathcal{E}^{MCS}]}}{\mathbb{E}_p[{p}_\mathcal{E}^{MCS}]}=\sqrt{\frac{1-p_\mathcal{E}}{Np_\mathcal{E}}}.
\end{equation}
This can be approximated by replacing $p_\mathcal{E}$ by the estimate $\widehat{p}_\mathcal{E}^{MCS}=\widehat{N}_\mathcal{E}/N$ for a given set of $N$ samples $\{\widehat{U}_1,\ldots,\widehat{U}_N\}$:
\begin{equation}\label{approx cov}
\delta({p}_\mathcal{E}^{MCS}|p_\mathcal{E},N)\approx\sqrt{\frac{1-\widehat{p}_\mathcal{E}^{MCS}}{N\widehat{p}_\mathcal{E}^{MCS}}}\stackrel{\triangle}{=}\widehat{\delta}_N^{MCS}.
\end{equation}

For the Bayesian interpretation, the posterior c.o.v. for the stochastic variable $p_\mathcal{E}$, conditional on the set of $N$ samples, follows from (\ref{posterior}):
\begin{equation}
\begin{split}
&\delta(p_\mathcal{E}|\widehat{N}_\mathcal{E},N)=\frac{\sqrt{\mathrm{Var}[{p}_\mathcal{E}|\widehat{N}_\mathcal{E},N]}}{\mathbb{E}[{p}_\mathcal{E}|\widehat{N}_\mathcal{E},N]}\\
&=\frac{\sqrt{1-\frac{\widehat{N}_\mathcal{E}+1}{N+2}}}{\sqrt{(N+3)\left(\frac{\widehat{N}_\mathcal{E}+1}{N+2}\right)}}\longrightarrow\sqrt{\frac{1-\widehat{p}_\mathcal{E}^{MCS}}{N\widehat{p}_\mathcal{E}^{MCS}}}=\widehat{\delta}_N^{MCS},
\end{split}
\end{equation} 
as $N\rightarrow\infty$. Therefore, the same expression $\widehat{\delta}_N^{MCS}$ can be used to assess the accuracy of the MCS estimate, even though the two c.o.v.s have distinct interpretations. 

The approximation $\widehat{\delta}_N^{MCS}$ for the two c.o.v.s reveals both the main advantage of the standard MCS method and its main drawback.  
   The main strength of MCS, which makes it very robust, is that its accuracy does not depend on the geometry of the domain $\mathcal{E}\subset\mathbb{R}^D$ and its dimension $D$. As long as an algorithm for generating i.i.d. samples from $p(u)$ is available, MCS, unlike many other methods (e.g. numerical integration), does not suffer from the ``curse
of dimensionality.'' 
 Moreover, an irregular, or even fractal-like, shape of  $\mathcal{E}$ will not affect the accuracy of MCS.

On the other hand, the serious drawback of MCS is that this method is not computationally efficient in estimating the \textit{small probabilities} $p_\mathcal{E}$ corresponding to \textit{rare events}, where from (\ref{cov}),
\begin{equation}\label{cov2}
\delta({p}_\mathcal{E}^{MCS}|p_\mathcal{E},N)\approx \frac{1}{\sqrt{Np_\mathcal{E}}}.
\end{equation}
Therefore, to achieve a prescribed level of accuracy $\delta<1$, the required total number of samples  is $N=(p_\mathcal{E}\delta^2)^{-1}\gg1$. For each sampled excitation $U_i$, a system analysis --- usually computationally very intensive --- is required to compute the corresponding system trajectory $X_i$ and to check whether $U_i$ belongs to $\mathcal{E}$. This makes MCS excessively costly and  inapplicable for generating rare events and estimating their small probabilities. Nevertheless, essentially all sampling-based methods for estimation of rare event probability are either based on MCS (e.g. Importance Sampling) or have it as a part of the algorithm (e.g. Subset Simulation).

\section{Importance Sampling}
The \textit{Importance Sampling} (IS) method belongs to the class of \textit{variance reduction techniques}  that aim to increase the accuracy of the estimates by constructing (sometimes biased) estimators with a smaller variance \cite{Asmusen,Dunn}. It seems it was first proposed in \cite{Kahn}, soon after the standard MCS method appeared.  

The inefficiency of MCS for rare event estimation stems from the fact that most of the generated samples $U_i\sim p(u)$ do not belong to $\mathcal{E}$ so that the vast majority
of the terms in the sum (\ref{MCSestimate}) are zero and only very few (if any) are equal to one. The basic idea of IS is to make use of the information available about the rare event $\mathcal{E}$ to generate samples that lie more frequently in $\mathcal{E}$ or in the \textit{important region} $\tilde{\mathcal{E}}\subset\mathcal{E}$ that accounts for most of the probability content in (\ref{pF}). Rather than estimating 
$p_\mathcal{E}$ as an average of many 0's and very few 1's like in $\widehat{p}_\mathcal{E}^{MCS}$, IS seeks to reduce the variance by constructing an estimator of the form ${p}_\mathcal{E}^{IS}=\frac{1}{N}\sum_{i=1}^{N'}w_i$, where $N'$ is an appreciable fraction of $N$ and the $w_i$ are small but not zero, ideally of the same order as the target probability, $w_i\approx p_\mathcal{E}$.

Specifically, for an appropriate PDF $q(u)$ on the excitation space $\mathbb{R}^D$, the integral in (\ref{pF=E[I]}) can be re-written as follows:
\begin{equation}\label{ISrewriting}
\begin{split}
p_\mathcal{E}&=\int_{\mathbb{R}^{D}}{I_\mathcal{E}(u)p(u)du}\\
&=\int_{\mathbb{R}^{D}}{\frac{I_\mathcal{E}(u)p(u)}{q(u)}q(u)du}=\mathbb{E}_q\left[\frac{I_\mathcal{E}p}{q}\right].
\end{split}
\end{equation}
The IS estimator is now constructed similarly to (\ref{MCSestimate}) by utilizing the law of large numbers:
\begin{equation}\label{ISestimate}
\begin{split}
p_\mathcal{E}\approx{p}_\mathcal{E}^{IS}&=\frac{1}{N}\sum_{i=1}^N\frac{I_{\mathcal{E}}(U_i)p(U_i)}{q(U_i)}\\
&=\frac{1}{N}\sum_{i=1}^NI_{\mathcal{E}}(U_i)w(U_i),
\end{split}
\end{equation}
where $U_1,\ldots,U_N$ are i.i.d. samples from $q(u)$, called the \textit{importance sampling density} (ISD), and $w(U_i)=\frac{p(U_i)}{q(U_i)}$ is the \textit{importance weight} of sample $U_i$. 

The IS estimator ${p}_\mathcal{E}^{IS}$ converges almost surely as $N\rightarrow\infty$ to $p_\mathcal{E}$ by the strong law of large numbers, provided that the support of $q(u)$, i.e. the domain in $\mathbb{R}^D$ where $q(u)>0$, contains the support of $I_\mathcal{E}(u)p(u)$. Intuitively, the latter condition guarantees that all points of $\mathcal{E}$ that can be generated by sampling from the original PDF $p(u)$, can also be generated by sampling from the ISD $q(u)$. Note that if $q(u)=p(u)$, then $w(U_i)=1$ and IS simply reduces to MCS, ${p}_\mathcal{E}^{MCS}={p}_\mathcal{E}^{IS}$. By choosing the ISD $q(u)$ appropriately, IS aims to obtain an estimator with a smaller variance.  

The IS estimator ${p}_\mathcal{E}^{IS}$ is also \textit{unbiased} with mean and variance:
\begin{equation}\label{mean var IS}
\begin{split}
\mathbb{E}_q[{p}_\mathcal{E}^{IS}]&=\mathbb{E}_q\left[\frac{1}{N}\sum_{i=1}^NI_\mathcal{E}(U_i)w(U_i)\right]\\
&=
\frac{1}{N}\sum_{i=1}^N\mathbb{E}_q\left[\frac{I_\mathcal{E}p}{q}\right]=p_\mathcal{E},\\
\mathrm{Var}_q[{p}_\mathcal{E}^{IS}]&=\frac{1}{N^2}\sum_{i=1}^N\mathrm{Var}_q\left[\frac{I_\mathcal{E}p}{q}\right]\\
&=\frac{1}{N}\left(\mathbb{E}_q\left[\frac{I_\mathcal{E}p^2}{q^2}\right]-p_\mathcal{E}^2\right).
\end{split}
\end{equation}
The IS method is summarized in the following pseudo-code.\\

\vspace{0mm} 
\hrule height 0.6pt \rule{0pt}{4mm}\centerline
{\textbf{Importance Sampling}}\rule{0pt}{4mm}
\hrule
\vspace{1mm}
\texttt{Input:}

\hspace{0.5cm}$\vartriangleright$ $N$, total number of samples.

\hspace{0.5cm}$\vartriangleright$ $q(u)$, importance sampling density.

\texttt{Algorithm:}

\hspace{0.5cm} Set $j=0$, counter for the number of samples in $\mathcal{E}$.

\hspace{0.5cm} \textbf{for} $i=1,\ldots,N$ \textbf{do}

\hspace{1.2cm}  Sample the input excitation  

\hspace{1.2cm} $U_i=(U_i(0),\ldots,U_i(T))\sim q(u)$.

\hspace{1.2cm} Compute the system trajectory 

\hspace{1.2cm} $X_i=(X_i(0),\ldots,X_i(T))$ 

\hspace{1.2cm} using the system model (\ref{model}) with $U(t)=U_i(t)$.

\hspace{1.2cm} \textbf{if}  $\max\limits_{t=0,\ldots,T}g(X_i(t))> b$

\hspace{1.6cm} $j\leftarrow j+1$

\hspace{1.6cm} Compute the importance weight of the $j^{th}$ 

\hspace{1.6cm}sample in $\mathcal{E}$, $w_j=\frac{p(U_i)}{q(U_i)}$. 

\hspace{1.2cm} \textbf{end if}

\hspace{0.5cm} \textbf{end for}

\hspace{0.5cm} $N_\mathcal{E}=j$, the total number of trials where the event

\hspace{0.5cm} $\mathcal{E}$ occurred.

\texttt{Output:}

\hspace{0.5cm}$\blacktriangleright$  $\widehat{p}_\mathcal{E}^{IS}=\frac{\sum_{j=1}^{N_\mathcal{E}}w_j}{N}$, IS estimate of $p_\mathcal{E}$.

\vspace{1mm}
\hrule
\vspace{5mm}

The most important task in applying IS for estimating small probabilities of rare events is the construction of the ISD, since the accuracy of $\widehat{p}_\mathcal{E}^{IS}$ depends critically on $q(u)$. If the ISD is ``good'', then one can get great improvement in efficiency over standard MCS. If, however, the ISD is chosen inappropriately so that for instance  $N_\mathcal{E}=0$ or the importance weights 
have a large variation, then IS will yield a very poor estimate. Both scenarios are demonstrated below in the Section ``Illustrative Example''.

It is straightforward to show that the \textit{optimal} ISD, which minimizes the variance in (\ref{mean var IS}), is simply the original PDF $p(u)$ conditional on the domain $\mathcal{E}$: 
\begin{equation}\label{OptimalISD}
q_{0}(u)=p(u|\mathcal{E})=\frac{I_\mathcal{E}(u)p(u)}{p_{\mathcal{E}}}.
\end{equation}
Indeed, in this case, all generated sample excitations satisfy   $U_i\in\mathcal{E}$, so their importance weights $w(U_i)=p_\mathcal{E}$, and the IS estimate  $\widehat{p}_\mathcal{E}^{IS}=p_\mathcal{E}$. Moreover, just one sample ($N=1$) generated from $q_0(u)$ is enough to find the probability $p_\mathcal{E}$ exactly.  Note, however, that this is a purely theoretical result since in practice sampling from the conditional distribution $p(u|\mathcal{E})$ is challenging, and, most importantly, it is impossible to compute $q_0(u)$: this would require the knowledge of $p_\mathcal{E}$, which is unknown. Nevertheless, this result indicates that the ISD $q(u)$ should be chosen as close to $q_0(u)$ as possible. In particular, most of the probability mass of $q(u)$ should be concentrated on $\mathcal{E}$. Based on these considerations, several ad hoc techniques for constructing ISDs have been developed, e.g.  variance scaling and mean shifting \cite{Bucklew}. 

In the special case of linear dynamics and Gaussian excitation, an extremely efficient algorithm for estimating the rare-event probability $p_\mathcal{E}$ in (\ref{pF}), referred to as ISEE (\textit{Importance Sampling using Elementary Events}), has been presented  \cite{AuBeck2001}. The choice of the ISD exploits known information about each elementary event, defined as an outcrossing of the performance threshold $b$ in  (\ref{F})  at a specific time $t\in\{0,\ldots,T\}$. The c.o.v. of the ISEE estimator for $N$ samples of $U$ from $p(u)$ is given by
\begin{equation}
\delta_N^{ISEE}=\frac{\alpha}{\sqrt{N}},
\end{equation}
where the proportionality constant $\alpha$ is close to $1$, regardless of how small the value of $p_\mathcal{E}$. In fact, $\alpha$ decreases slightly as $p_\mathcal{E}$ decreases, exhibiting the opposite behavior to MCS. 

In general, it is known that in many practical cases of rare event estimation it is difficult to construct a good ISD that leads to a low-variance IS estimator, especially if the dimension of the uncertain excitation space $\mathbb{R}^D$ is large, as it is in dynamic reliability problems  \cite{IShighdim}. A geometric explanation as to why IS in often inefficient in high dimensions is given in \cite{KatZuev}.  Au \cite{Au}  has presented an efficient IS method for estimating $p_\mathcal{E}$ in (\ref{pF}) for elasto-plastic systems subject to Gaussian excitation. In recent years, substantial progress has been made by tailoring the \textit{sequential importance sampling} (SIS) methods \cite{Liu}, where the ISD is iteratively refined, to rare event problems. SIS and its modifications have been successfully used for estimating rare events in
 dynamic portfolio credit risk \cite{Deng}, structural reliability \cite{ALIS}, and other areas.

\section{Subset Simulation}

The \textit{Subset Simulation} (SS) method \cite{SS} is an advanced stochastic simulation method for estimating rare events which is based on \textit{Markov chain Monte Carlo} (MCMC) \cite{Liu,Robert}. The basic idea behind SS is to represent a very small probability $p_\mathcal{E}$ of the rare event $\mathcal{E}$ as a product of larger probabilities of ``more-frequent'' events and then estimate these larger probabilities  separately.  To implement this idea, let
\begin{equation}\label{filtration}
\mathbb{R}^D\equiv\mathcal{E}_0\supset\mathcal{E}_1\ldots\supset\mathcal{E}_L\equiv\mathcal{E}
\end{equation}
be a sequence of nested subsets of the uncertain excitation space starting  from the entire space $\mathcal{E}_0=\mathbb{R}^D$ and shrinking to the target rare event $\mathcal{E}_L=\mathcal{E}$. By analogy with (\ref{F}), subsets $\mathcal{E}_i$ can be defined by relaxing the value of the critical threshold $b$:
\begin{equation}\label{subsets}
\mathcal{E}_i=\left\{U\in\mathbb{R}^D : \max\limits_{t=0,\ldots,T}g(X(t))> b_i\right\},
\end{equation}
where $b_1<\ldots<b_L=b$. In the actual implementation of SS, the number of subsets $L$ and the values of intermediate thresholds $\{b_i\}$ are chosen adaptively. 

 Using the notion of conditional probability and exploiting the nesting of the subsets, the target probability $p_\mathcal{E}$ can be factorized as follows:
\begin{equation}\label{product}
p_\mathcal{E}=\prod_{i=1}^L\mathbb{P}(\mathcal{E}_i|\mathcal{E}_{i-1}).
\end{equation}
An important observation is that by choosing the intermediate thresholds $\{b_i\}$ appropriately, the conditional events $\{\mathcal{E}_i|\mathcal{E}_{i-1}\}$ can be made more frequent, and their probabilities can be made large enough to be amenable to efficient estimation by MCS-like methods. 

The first probability $\mathbb{P}(\mathcal{E}_1|\mathcal{E}_0)=\mathbb{P}(\mathcal{E}_1)$ can be readily estimated by standard MCS:
\begin{equation}\label{p1}
\mathbb{P}(\mathcal{E}_1)\approx \frac{1}{n}\sum_{j=1}^nI_{\mathcal{E}_1}(U_j),
\end{equation}
where $U_1,\ldots,U_n$ are i.i.d. samples from $p(u)$. 
 Estimating the remaining probabilities $\mathbb{P}(\mathcal{E}_i|\mathcal{E}_{i-1})$, $i\geq2$, is more challenging since one needs to generate samples from the conditional distribution $p(u|\mathcal{E}_{i-1})=\frac{I_{\mathcal{E}_{i-1}}(u)p(u)}{\mathbb{P}(\mathcal{E}_{i-1})}$, which, in general, is not a trivial task. Notice that a sample $U$ from $p(u|\mathcal{E}_{i-1})$ is one drawn from $p(u)$ that lies in $\mathcal{E}_{i-1}$. However, it is not efficient to use MCS for generating samples from $p(u|\mathcal{E}_{i-1})$: sampling from $p(u)$ and accepting only those samples that belong to $\mathcal{E}_{i-1}$ is computationally very expensive, especially at higher levels~$i$.

In standard SS, samples from the conditional distribution $p(u|\mathcal{E}_{i-1})$ are generated by the \textit{modified Metropolis algorithm} (MMA) \cite{SS} which belongs to the family of MCMC methods for sampling from complex probability distributions that are difficult to sample
from directly \cite{Liu,Robert}. An alternative strategy --- \textit{splitting} --- is described in the next section.  

The MMA algorithm is a component-wise version of the original Metropolis algorithm \cite{MA}. It is specifically tailored for sampling from high-dimensional conditional distributions and works as follows. First, without loss of generality, assume that $p(u)=\prod_{k=1}^Dp_k(u_k)$, i.e. components of $U$ are independent. This assumption is indeed not a limitation, since in simulation
one always starts from independent variables to generate correlated excitation histories $U$. Suppose further, that some vector $U_1\in\mathbb{R}^D$ is already distributed according to the target conditional distribution,  $U_1\sim p(u|\mathcal{E}_{i-1})$. MMA prescribes how to generate another vector $U_2\sim p(u|\mathcal{E}_{i-1})$ and it consists of two steps:
\begin{enumerate}
	\item Generate a ``candidate'' state $V$ as follows: first, for each component $k=1,\ldots,D$ of $V$, sample $\nu(k)$ from   the symmetric univariate \textit{proposal distribution} $q_{k,i}(\nu|U_1(k))$ centered on the $k^{\mathrm{th}}$ component of $U_1$, where symmetry means that $q_{k,i}(\nu|u)=q_{k,i}(u|\nu)$; then, compute the \textit{acceptance ratio} $r_k=\frac{p_k(\nu(k))}{p_k(U_1(k))}$; finally, set
	\begin{equation}\label{V(k)}
	V(k)=\left\{
	\begin{array}{ll}
	\nu(k), & \hbox{with prob. } \min\{1,r_k\}, \\
	U_1(k), & \hbox{with prob. } 1-\min\{1,r_k\}.
	\end{array}
	\right.
	\end{equation}
	\item Accept or reject the candidate state $V$:
	\begin{equation}\label{U2}
	U_2=\left\{
	\begin{array}{ll}
	V, & \hbox{if } V\in \mathcal{E}_{i-1}, \\
	U_1, & \hbox{if } V\notin\mathcal{E}_{i-1}.
	\end{array}
	\right.
	\end{equation}
\end{enumerate}

It can be shown that $U_2$ generated by MMA is indeed distributed according to the target conditional distribution $p(u|\mathcal{E}_{i-1})$ when $U_1$ is \cite{SS}. For a detailed discussion of MMA, the reader is referred to \cite{BSS}. 

The procedure for generating conditional samples at level $i$ is as follows. Starting from a ``seed'' $U_1~\sim~p(u|\mathcal{E}_{i-1})$, one can now use MMA to generate a sequence of random vectors $U_1,\ldots,U_n$, called a \textit{Markov chain}, distributed according to $p(u|\mathcal{E}_{i-1})$. At each step, $U_{j}$ is used to generate the next state $U_{j+1}$. Note that although these MCMC samples are identically distributed, they are clearly not independent: the correlation between successive samples is due to the proposal PDFs $\{q_{k,i}\}$ at level $i$ that govern the generation of  $U_{j+1}$  from  $U_j$. Nevertheless, $U_1,\ldots,U_n$ can still be used for statistical averaging as if they were i.i.d, although with certain reduction in efficiency \cite{SS}. In particular, similarly to (\ref{p1}), the conditional probability $\mathbb{P}(\mathcal{E}_i|\mathcal{E}_{i-1})$, can be estimated as follows:
\begin{equation}\label{pi}
\mathbb{P}(\mathcal{E}_i|\mathcal{E}_{i-1})\approx\frac{1}{n}\sum_{j=1}^nI_{\mathcal{E}_i}(U_j).
\end{equation}

To obtain an estimator for the target probability $p_\mathcal{E}$, it remains to multiply the MCS (\ref{p1}) and MCMC (\ref{pi}) estimators of all factors in (\ref{product}). In real applications, however, it is often difficult to rationally define the subsets $\{\mathcal{E}_i\}$ in advance, since it is not clear how to specify the values of the intermediate thresholds $\{b_i\}$. In SS, this is done adaptively. Specifically, let $U_1^{(0)},\ldots,U_n^{(0)}$ be the MCS samples from $p(u)$, $X_1^{(0)},\ldots,X_n^{(0)}$ be the corresponding trajectories from (\ref{model}), and $G_j^{(0)}=\max_{t=0,\ldots,T}g(X_j^{(0)}(t))$ be the resulting performance values. Assume that the sequence $\{G_j^{(0)}\}$ is ordered in non-increasing order, i.e. $G_1^{(0)}\geq\ldots\geq G_n^{(0)}$, renumbering the samples where necessary. Define the first intermediate threshold $b_1$ as follows:
\begin{equation}
b_1=\frac{G^{(0)}_{np_0}+G^{(0)}_{np_0+1}}{2},
\end{equation}
where $p_0$ is a chosen probability satisfying $0<p_0<1$. This choice of $b_1$ has two immediate consequences: first, the MCS estimate of $\mathbb{P}(\mathcal{E}_1)$ in (\ref{p1}) is exactly $p_0$, and, second, $U_1^{(0)}\ldots,U^{(0)}_{np_0}$ not only belong to $\mathcal{E}_1$, but also are distributed according to the conditional distribution $p(u|\mathcal{E}_1)$. Each of these $np_0$ samples can now be used as mother seeds in MMA to generate $(\frac{1}{p_0}-1)$ offspring, giving a total of  $n$ samples $U_1^{(1)},\ldots,U_n^{(1)}\sim p(u|\mathcal{E}_1)$. Since these seeds start in the stationary state $p(u|\mathcal{E}_1)$ of the Markov chain, this MCMC method gives \textit{perfect sampling}, i.e. no wasteful burn-in period is needed.  Similarly, $b_2$ is defined as 
\begin{equation}
b_2=\frac{G^{(1)}_{np_0}+G^{(1)}_{np_0+1}}{2},
\end{equation}
where $\{G_j^{(1)}\}$ are the (ordered) performance values corresponding to excitations $\{U_j^{(1)}\}$. Again  by construction, the estimate (\ref{pi}) gives $\mathbb{P}(\mathcal{E}_2|\mathcal{E}_1)\approx p_0$, and $U_1^{(1)},\ldots,U_{np_0}^{(1)}\sim p(u|\mathcal{E}_2)$. The SS method proceeds in this manner until the target rare event $\mathcal{E}$ is reached and is sufficiently sampled. All but the last factor in (\ref{product}) are approximated by $p_0$, and the last factor $\mathbb{P}(\mathcal{E}|\mathcal{E}_{L-1})\approx\frac{n_\mathcal{E}}{n}\geq p_0$, where $n_\mathcal{E}$ is the number of samples in $\mathcal{E}$ among $U_1^{(L-1)},\ldots,U_n^{(L-1)}\sim p(u|\mathcal{E}_{L-1})$. The method is more formally summarized in the following pseudo-code. 

\vspace{2mm} \hrule height 0.6pt \rule{0pt}{4mm}\centerline
{\textbf{Subset Simulation}}\vspace{.5mm}\rule{0pt}{4mm}
\hrule
\vspace{1mm}
\texttt{Input:}

\hspace{0.5cm}$\vartriangleright$ $n$, number of samples per
conditional level.

\hspace{0.5cm}$\vartriangleright$ $p_0$, level 
probability; e.g. $p_0=0.1$.

\hspace{0.5cm}$\vartriangleright$ $\{q_{k,i}\}$, proposal distributions; 

\hspace{0.8cm} e.g. $q_{k,i}(\nu|u)=\mathcal{N}(\nu|u,\sigma^2_{k,i})$. 

\texttt{Algorithm:}

\hspace{0.5cm} Set $i=0$, number of conditional level.

\hspace{0.5cm} Set $n_\mathcal{E}^{(0)}=0$, number of the MCS samples in $\mathcal{E}$.

\hspace{0.5cm} Sample the input excitations
$U_1^{(0)},\ldots, U_n^{(0)}\sim p(u)$.

\hspace{0.5cm} Compute the corresponding  trajectories

\hspace{0.5cm} $X_1^{(0)},\ldots, X_n^{(0)}$.

\hspace{0.5cm} \textbf{for} $j=1,\ldots,n$ \textbf{do}

\hspace{1.2cm} \textbf{if} $G^{(0)}_j=\max\limits_{t=0,\ldots,T}g(X^{(0)}_j(t))>b$ \textbf{do}

\hspace{1.6cm} $n_\mathcal{E}^{(0)}\leftarrow n_\mathcal{E}^{(0)}+1$

\hspace{1.2cm} \textbf{end if}

\hspace{0.5cm} \textbf{end for}

\hspace{0.5cm} \textbf{while} $n_\mathcal{E}^{(i)}/n<p_0$ \textbf{do}

\hspace{1.7cm} $i\leftarrow i+1$, a new subset $\mathcal{E}_i$ is needed.

\hspace{1.7cm} Sort $\{U_j^{(i-1)}\}$ so that 

\hspace{1.7cm} $G_{1}^{(i-1)}\geq
G_{2}^{(i-1)}\geq\ldots\geq G_{n}^{(i-1)}$.

\hspace{1.7cm} Define the $i^{\mathrm{th}}$ intermediate threshold: 

\hspace{1.7cm} $b_i=\left(G_{{np_0}}^{(i-1)}+G_{{np_0+1}}^{(i-1)}\right)\left/2\right.$.

\hspace{1.7cm} \textbf{for} $j=1,\ldots,np_0$ \textbf{do}

\hspace{2.4cm} Using
$W_{j,1}=U^{(i-1)}_{j}\sim p(u|\mathcal{E}_i)$ as a 

\hspace{2.4cm} seed, use MMA to generate $(\frac{1}{p_0}-1)$ 

\hspace{2.4cm}  additional states of a Markov chain

\hspace{2.4cm} $W_{j,1},\ldots,W_{j,1/p_0}\sim p(u|\mathcal{E}_i)$.

\hspace{1.7cm} \textbf{end for}

\hspace{1.7cm} Renumber:

\hspace{1.7cm} $\{W_{j,s}\}_{j=1,s=1}^{np_0,1/p_0} \mapsto
U_1^{(i)},\ldots,U_n^{(i)}\sim p(u|\mathcal{E}_i)$.

\hspace{1.7cm}  Compute the corresponding  trajectories

\hspace{1.7cm} $X_1^{(i)},\ldots, X_n^{(i)}$.

\hspace{1.7cm} \textbf{for} $j=1,\ldots,n$ \textbf{do}

\hspace{2.4cm} \textbf{if} $G^{(i)}_j=\max\limits_{t=0,\ldots,T}g(X^{(i)}_j(t))>b$ \textbf{do}

\hspace{2.8cm} $n_\mathcal{E}^{(i)}\leftarrow n_\mathcal{E}^{(i)}+1$

\hspace{2.4cm} \textbf{end if}

\hspace{1.7cm} \textbf{end for}

\hspace{0.5cm} \textbf{end while}

\hspace{0.5cm}  $L=i+1$, number of levels, i.e. subsets  $\mathcal{E}_i$ in  (\ref{filtration})

\hspace{0.5cm}   and (\ref{subsets}).

\hspace{0.5cm}  $N=n+n(1-p_0)(L-1)$, total number of samples.

\texttt{Output:}

\hspace{0.5cm}$\blacktriangleright$ $\widehat{p}_\mathcal{E}^{SS}=p_0^{L-1}\frac{n_\mathcal{E}^{(L-1)}}{n}$, SS estimate of $p_F$.
\vspace{1mm}
\hrule
\vspace{5mm}
  
Implementation details of SS, in particular the choice of level probability $p_0$ and proposal distributions $\{q_k\}$, are thoroughly discussed in \cite{BSS}. It has been confirmed that $p_0=0.1$ proposed in the original paper \cite{SS} is a nearly optimal value. The choice of $\{q_{k,i}\}$ is more delicate, since the efficiency of MMA strongly depends on the proposal PDF variances in a non-trivial way: proposal PDFs with both small and large variance  tend to increase the correlation between successive samples, making statistical averaging in (\ref{pi}) less efficient. In general, finding the optimal variance of proposal distributions is a challenging task  not only for MMA, but also for almost all MCMC algorithms. Nevertheless,  it has been found in many applications that using $q_{k,i}(\nu|u)=\mathcal{N}(\nu|u,\sigma^2_{k,i})$,  the Gaussian distribution with mean $u$ and variance $\sigma^2_{k,i}$,  yields good efficiency if $\sigma^2_{k,i}=\sigma_0^2$ and $p(u)$ is a multi-dimensional Gaussian with all variances equal to $\sigma_0^2$. For an adaptive strategy for choosing $\{q_{k,i}\}$, the reader is referred to \cite{BSS}; for example, $\sigma^2_{k,i}=\sigma^2_{i}$ can be chosen so that the observed average acceptance rate in MMA, based on a subset of samples at level $i$, lies in the interval $[0.3,0.5]$.

It can be shown \cite{SS,SSbook} that, given $p_\mathcal{E}$, $p_0$, and the total number of samples $N$, the c.o.v. of the SS estimator ${p}_\mathcal{E}^{SS}$ is given by
\begin{equation}\label{covSS}
\delta^2({p}_\mathcal{E}^{SS}|p_\mathcal{E},p_0,N)=\frac{(1+\gamma)(1-p_0)}{Np_0(\ln p_0^{-1})^r}(\ln p_\mathcal{E}^{-1})^r,
\end{equation}
where $2\leq r\leq3$ and $\gamma$ is approximately a constant that depends on the state correlation of the Markov chain at each level. Numerical experiments show that $r=2$ gives a good approximation to the c.o.v. and that $\gamma\approx3$ if the proposal variance $\sigma_i^2$ for each level is appropriately chosen \cite{SS,SSbook,BSS}. It follows from (\ref{cov2}) that  $\delta^2_{MCS}\propto p_\mathcal{E}^{-1}$ for MCS,  while for SS, $\delta^2_{SS}\propto (\ln p_\mathcal{E}^{-1})^r$. This drastically different scaling behavior of the c.o.v.'s with small $p_\mathcal{E}$ directly exhibits the improvement in efficiency. 

To compare an advanced stochastic simulation algorithm directly with MCS, which is always applicable (but not efficient) for rare event estimation, \cite{BA05} introduced the relative computation efficiency of an algorithm, $\eta_A$, which is defined as the ratio of the number of samples $N_{MCS}$ required by MCS to the number of samples $N_A$ required by the algorithm for the same c.o.v. $\delta$. The \textit{relative efficiency} of SS is then
\begin{equation}
\begin{split}
\eta_{SS}&=\frac{N_{MCS}}{N_{SS}}=\frac{p_0(\ln p_0^{-1})^r}{(1+\gamma)(1-p_0)p_\mathcal{E}(\ln p_\mathcal{E}^{-1})^r}\\
&\approx\frac{0.03p_\mathcal{E}^{-1}}{(\log_{10} p_\mathcal{E}^{-1})^2},
\end{split}
\end{equation}
for $r=2$, $\gamma=3$, and $p_0=0.1$.
For rare events, $p_\mathcal{E}^{-1}$ is very large, and, as expected, SS outperforms MCS; for example, if $p_\mathcal{E}=10^{-6}$, then $\eta_{SS}\approx800$. 

In recent years, a number of modifications of SS have been proposed, including SS with Splitting \cite{ChingAuBeck} (described in the next section), Hybrid SS \cite{ChingBeckAu}, Two-Stage SS \cite{KatafygiotisCheung}, Spherical SS \cite{KatafygiotisCheung2}, and SS with delayed rejection~\cite{MMHA}. A Bayesian post-processor for SS, which generalizes the Bayesian interpretation of MCS described above, was developed in \cite{BSS}.  In the original paper \cite{SS}, SS was developed  for estimating  reliability of complex civil engineering structures such as tall buildings and bridges at risk from earthquakes. It was applied for this purpose in \cite{IShighdim} and \cite{JB08}. SS and its modifications have also been successfully applied to rare event simulation in fire risk analysis \cite{fire}, aerospace  \cite{aerospace,Thu07},  nuclear  \cite{Cad12}, wind  \cite{wind} and geotechnical engineering \cite{San11}, and other fields. A detailed exposition  of SS on an introductory level and a MATLAB code implementing the above pseudo-code is given in \cite{ZuevSS}. For more advanced and complete reading, the fundamental monograph on SS \cite{SSbook} is strongly recommended.

\section{Splitting}
In the previously presented stochastic simulation methods, samples of the input and output discrete-time histories, $\{U(t): t=0,\ldots,T\}\subset\mathbb{R}^m$ and $\{X(t): t=0,\ldots,T\}\subset\mathbb{R}^n$, are viewed geometrically as vectors $U$ and $X$  that define points in the vector spaces $\mathbb{R}^{(T+1)m}$ and $\mathbb{R}^{(T+1)n}$, respectively. In the splitting method, however, samples of the input and output histories are viewed as trajectories defining paths of length $(T+1)$ in $\mathbb{R}^m$ and  $\mathbb{R}^n$, respectively. Samples that reach a certain designated subset in the input or output spaces at some time are treated as ``mothers'' and are then split into multiple offspring trajectories by separate sampling of the input histories subsequent to the splitting time. These multiple trajectories can themselves subsequently be treated as mothers if they reach another designated subset nested inside the first subset at some later time, and so be split into multiple offspring trajectories. This is continued until a certain number of the trajectories reach the smallest nested subset corresponding to the rare event of interest.

Splitting methods were originally introduced by Kahn and Harris  \cite{Kahn51} and they have been extensively studied (for example, \cite{ChingAuBeck,PSMM94,VA02,BK12}). We describe splitting here by using the framework of Subset Simulation where the only change is that the conditional sampling in the nested subsets is done by splitting the trajectories that reach each subset, rather than using them as seeds to generate more samples from Markov chains in their stationary state. As a result, only standard Monte Carlo simulation is needed, instead of MCMC simulation. 

The procedure in \cite{ChingAuBeck} is followed here to generate offspring trajectories at the $i^{\mathrm{th}}$ level $(i=1,\ldots,L)$ of Subset Simulation from each of the mother trajectories in $\mathcal{E}_i$ constructed from samples from the previous level, except that we present it from the viewpoint of trajectories in the input space, rather than the output space. Therefore, at the $i^{\mathrm{th}}$ level, each of the $np_0$ sampled input histories $U_j$, $j=1,\ldots,np_0$, from the previous level that satisfy $U_j\in\mathcal{E}_i$, as defined in (\ref{subsets}) (so the corresponding output history $X_j$ satisfies $\max\limits_{t=0,\ldots,T}g(X_j(t))> b_i$), are split at their first-passage time 
\begin{equation}
t_j=\min\{t=0,\ldots,T: g(X_j(t))> b_i\}
\end{equation}
This means that the mother trajectory $U_j$ is partitioned as $[U_j^-,U_j^+]$ where $U_j^-=[U_j(0),\ldots,U_j(t_j)]$ and $U_j^+=[U_j(t_j+1),\ldots,U_j(T)]$; then a subtrajectory  sample $\widetilde{U}_j^+=[\widetilde{U}_j(t_j+1),\ldots,\widetilde{U}_j(T)]$ is drawn from 
\begin{equation}
\begin{split}
p(u_j^+|U_j^-,\mathcal{E}_i)&=\frac{\mathbb{P}(\mathcal{E}_i|u_j^+,U_j^-)}{\mathbb{P}(\mathcal{E}_i|U_j^-)}p(u_j^+|U_j^-)\\
&=p(u_j^+|U_j^-)=p(u_j^+),
\end{split}
\end{equation}
where the last equation follows if one assumes independence of the $U_j(t), t=0,\ldots,T$ (although it is not necessary). Also, $\mathbb{P}(\mathcal{E}_i|u_j^+,U_j^-)=1=\mathbb{P}(\mathcal{E}_i|U_j^-)$. Note that the new input sample $\widetilde{U}_j=[U_j^-,\widetilde{U}_j^+]$ also lies in $\mathcal{E}_i$ since it has the subtrajectory $U_j^-$ in common with $U_j$, which implies that the corresponding outputs at the first-passage time $t_j$ are equal: $\widetilde{X}_j(t_j)=X_j(t_j)>b_i$. The offspring trajectory $\widetilde{U}_j$ is a sample from $p(u)$ lying in $\mathcal{E}_i$ and so, like its mother $U_j$, it is a sample from $p(u|\mathcal{E}_i)$. This process is repeated to generate $(\frac{1}{p_0}-1)$ such offspring trajectories from each mother trajectory, giving a total of $np_0(\frac{1}{p_0}-1)+np_0=n$ input histories that are samples from $p(u|\mathcal{E}_i)$ at the $i^{\mathrm{th}}$ level. 

The pseudo-code for the splitting version of Subset Simulation is the same as the previously presented pseudo-code for the MCMC version except that the part describing the generation of conditional samples at level $i$ using the MMA algorithm is  replaced by:

\vspace{3mm} \hrule height 0.6pt \rule{0pt}{4mm}\centerline
{\textbf{Generation of conditional samples}} \\ \centerline{\textbf{at level $i$ with Splitting}}\rule{0pt}{4mm}
\hrule
\vspace{2mm}
 \textbf{for} $j=1,\ldots,np_0$ \textbf{do}

\hspace{0.5cm} Using $U_j^{(i-1)}\sim p(u|\mathcal{E}_i)$ as a mother trajectory, 

\hspace{0.5cm} generate $(\frac{1}{p_0}-1)$ offspring trajectories by splitting

\hspace{0.5cm}  of this input trajectory.

 \textbf{end for}
\vspace{2mm}
\hrule
\vspace{5mm}

To generate the same number of samples $n$ at a level, the splitting version of Subset Simulation is slightly more efficient than the MCMC version using MMA because when generating the conditional samples, the input offspring trajectories $\widetilde{U}=[\widetilde{U}^-,\widetilde{U}^+]$ already have available the first part $\widetilde{X}^-$  of the corresponding output trajectory $\widetilde{X}=[\widetilde{X}^-,\widetilde{X}^+]$.
Thus, (\ref{model}) need only be solved for $\widetilde{X}^+$  starting from the final value of  $\widetilde{X}^-$ (which corresponds to the first-passage time of the trajectory). A disadvantage of the splitting version is that it cannot handle parameter uncertainty in the model in (\ref{model}) since the offspring trajectories must use (\ref{model}) with the same parameter values as their mothers. Furthermore, the splitting version applies only to dynamic problems, as considered here. The MCMC version of Subset Simulation can handle parameter uncertainty and is applicable to both static and dynamic uncertainty quantification problems.

Ching, Au and Beck  \cite{ChingAuBeck} discuss the statistical properties of the estimators corresponding to (\ref{p1}) and (\ref{pi}) when the sampling at each level is done by the trajectory splitting method. They show that as long as the conditional probability in Subset Simulation satisfies $p_0\geq0.1$, the coefficient of variation for $p_\mathcal{E}$ when estimating it by (\ref{product}) and (\ref{pi}) is insensitive to $p_0$.

Ching, Beck  and Au \cite{ChingBeckAu} also introduce a hybrid version of Subset Simulation that combines some advantages of the splitting and MCMC versions when generating the conditional samples $U_j, j=1,\ldots,n$ at each level. It is limited to dynamic problems because of the splitting but it can handle parameter uncertainty through using MCMC. All three variants of Subset Simulation are applied to a series of benchmark reliability problems in \cite{ACB07}; their results imply that for the same computational effort in the dynamic benchmark problems, the hybrid version gives slightly better accuracy for the rare-event probability than the MCMC version. For a comparison between these results and those of other stochastic simulation methods that are applied to some of the same benchmark problems (e.g. Spherical Subset Simulation, Auxiliary Domain Method and Line Sampling), the reader may wish to check \cite{SP07}. 

\section{Illustrative Example}\label{sec:Example}
To illustrate MCS, IS, and SS with MCMC and Splitting for rare event estimation, consider the following forced Lorenz system of ordinary differential equations:
\begin{align}\label{lorenz1}
\dot{X_1}&=\sigma(X_2-X_1)+U(t),\\
\dot{X_2}&=rX_1 - X_2 - X_1X_3,\label{lorenz2}\\
\dot{X_3}&=X_1X_2-bX_3, \label{lorenz3}
\end{align}
where $X(t)=(X_1(t),X_2(t),X_3(t))$ defines the system state at time $t$ and $U(t)$ is the external excitation to the system. If $U(t)\equiv0$,   these are the original equations due to E.~N.~Lorenz that he derived from a model of fluid convection \cite{Lorenz}. In this example, the three parameters $\sigma, r$, and $b$ are set to $\sigma=3$, $b=1$, and $r=26$. It is well-know (e.g. \cite{Sparrow}) that in this case, the Lorenz system has three unstable equilibrium points, one of which is
\begin{equation}\label{eq_point}
X^*=\left(\sqrt{b(r-1)},\sqrt{b(r-1)},r-1\right)=(5,5,25),
\end{equation}
that lies on one ``wing'' of the ``butterfly'' attractor. Let \begin{equation}\label{X(0)}
X(0)=X^*+(1/2,1/2,1/2)=(5.5,5.5,25.5)
\end{equation}
be the initial condition, and $X(t)$ be the corresponding solution. Lorenz showed \cite{Lorenz} that the solution of (\ref{lorenz1},\ref{lorenz2},\ref{lorenz3}) with  $U(t)\equiv0$ always (for any $t$) stays inside the bounding ellipsoid $\mathbb{E}$: 
\begin{equation}
\frac{X_1(t)^2}{R^2\frac{b}{\sigma}}+\frac{X_2(t)^2}{bR^2}+\frac{(X_3(t)-R)^2}{R^2}\leq 1, \hspace{3mm} R=r+\sigma
\end{equation}

\begin{figure}[t]
	\centerline{\includegraphics[angle=0,scale=0.45]{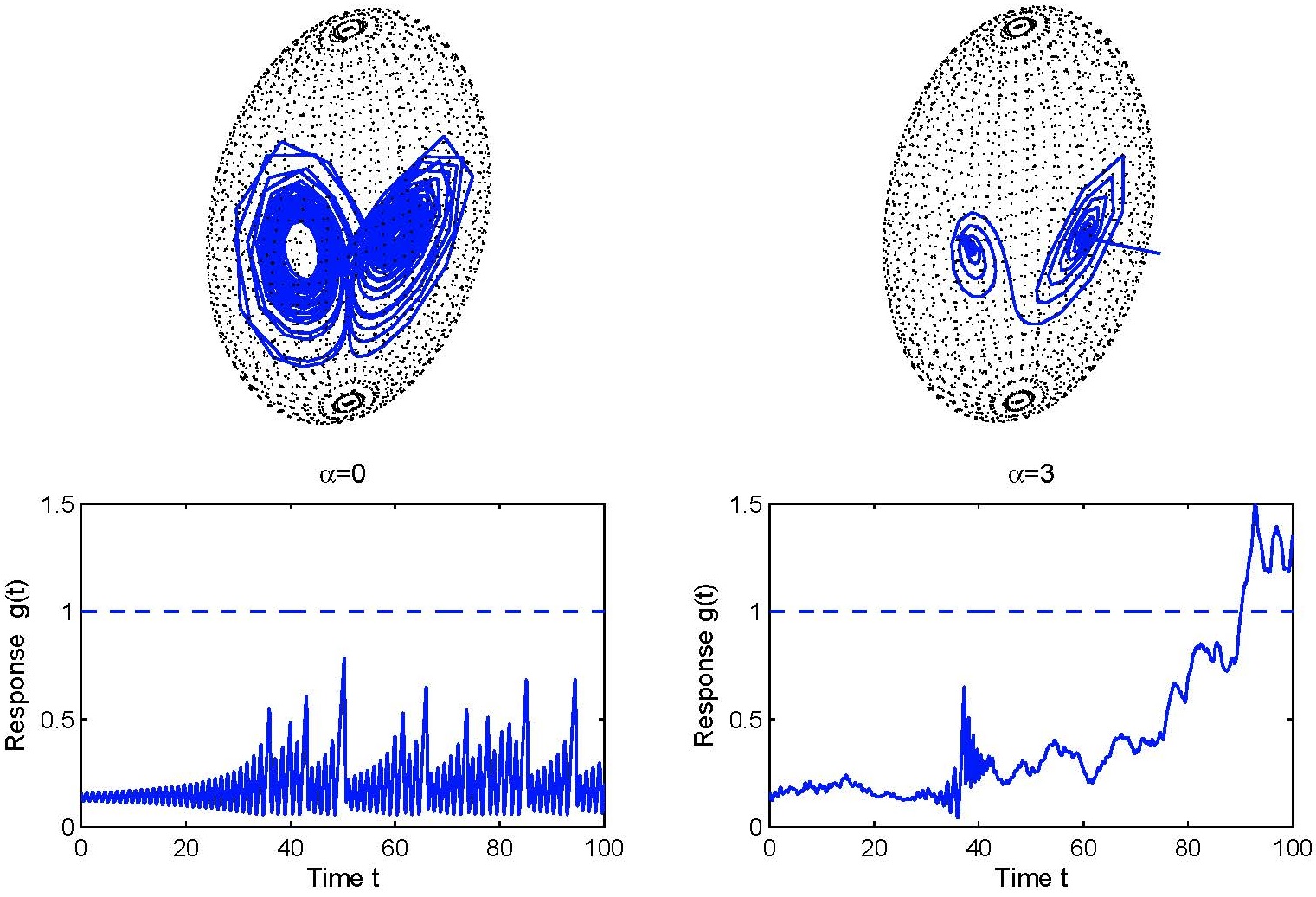}}
	\caption{\footnotesize The left column shows the solution of the unexcited Lorenz system ($\alpha=0$) enclosed in the bounding ellipsoid $\mathbb{E}$ (top) and the corresponding response function $g(t)$ (bottom), where $t\in[0,T]$, $T=100$. The right top panel shows the solution of the forced Lorenz system ($\alpha=3$) that corresponds to an excitation $U\in\mathcal{E}$. As it is clearly seen, this solution leaves the ellipsoid $\mathbb{E}$. According to the response function $g(t)$ shown in the right bottom panel, this first-passage event happens around $t=90$.} \label{fig1}
\end{figure}

Suppose that the system is now excited by $U(t)=\alpha B(t)$, where  $B(t)$ is the standard Brownian process (Gaussian white noise) and $\alpha$ is some scaling constant. The uncertain stochastic excitation $U(t)$ makes the corresponding system trajectory $X(t)$ also stochastic. Let us say that the event $\mathcal{E}$ occurs if $X(t)$ leaves the bounding ellipsoid $\mathbb{E}$ during the time interval of interest $[0,T]$. 

The discretization of the excitation $U$ is obtained by the standard discretization of the Brownian process:
\begin{equation}
\begin{split}
U(0)&=0, \\
U(k)&=\alpha B(k\Delta t)=U(k-1)+\alpha\sqrt{\Delta t}Z_k\\
&=\alpha\sqrt{\Delta t}\sum_{i=1}^k Z_i, 
\end{split}
\end{equation}
where $\Delta t=0.1$s is the sampling interval, $k=1,\ldots,D=T/\Delta t$, and $Z_1,\ldots, Z_{D}$ are i.i.d. standard Gaussian random variables. The target domain $\mathcal{E}\subset\mathbb{R}^D$ is then
\begin{equation}
\mathcal{E}=\{(Z_1,\ldots,Z_D) : \max_{0\leq k\leq D} g(k)>1\},
\end{equation}
where the system response $g(k)$ at time $t=k\Delta t$ is 
\begin{equation}
g(k)=\frac{X_1(k\Delta t)^2}{R^2\frac{b}{\sigma}}+\frac{X_2(k\Delta t)^2}{bR^2}+\frac{(X_3(k\Delta t)-R)^2}{R^2}.
\end{equation}

Figure~\ref{fig1} shows the solution of the unforced Lorenz system (with $\alpha=0$ so $U(t)=0$), and an example of the solution of the forced system (with $\alpha=3$) that corresponds to excitation $U\in\mathcal{E}$ (slightly abusing notation, $U=U(Z_1,\ldots,Z_D)\in\mathcal{E}$ means that the corresponding Gaussian vector $(Z_1,\ldots,Z_D)\in\mathcal{E}$).

\begin{figure}[t]
	\centerline{\includegraphics[angle=0,scale=0.45]{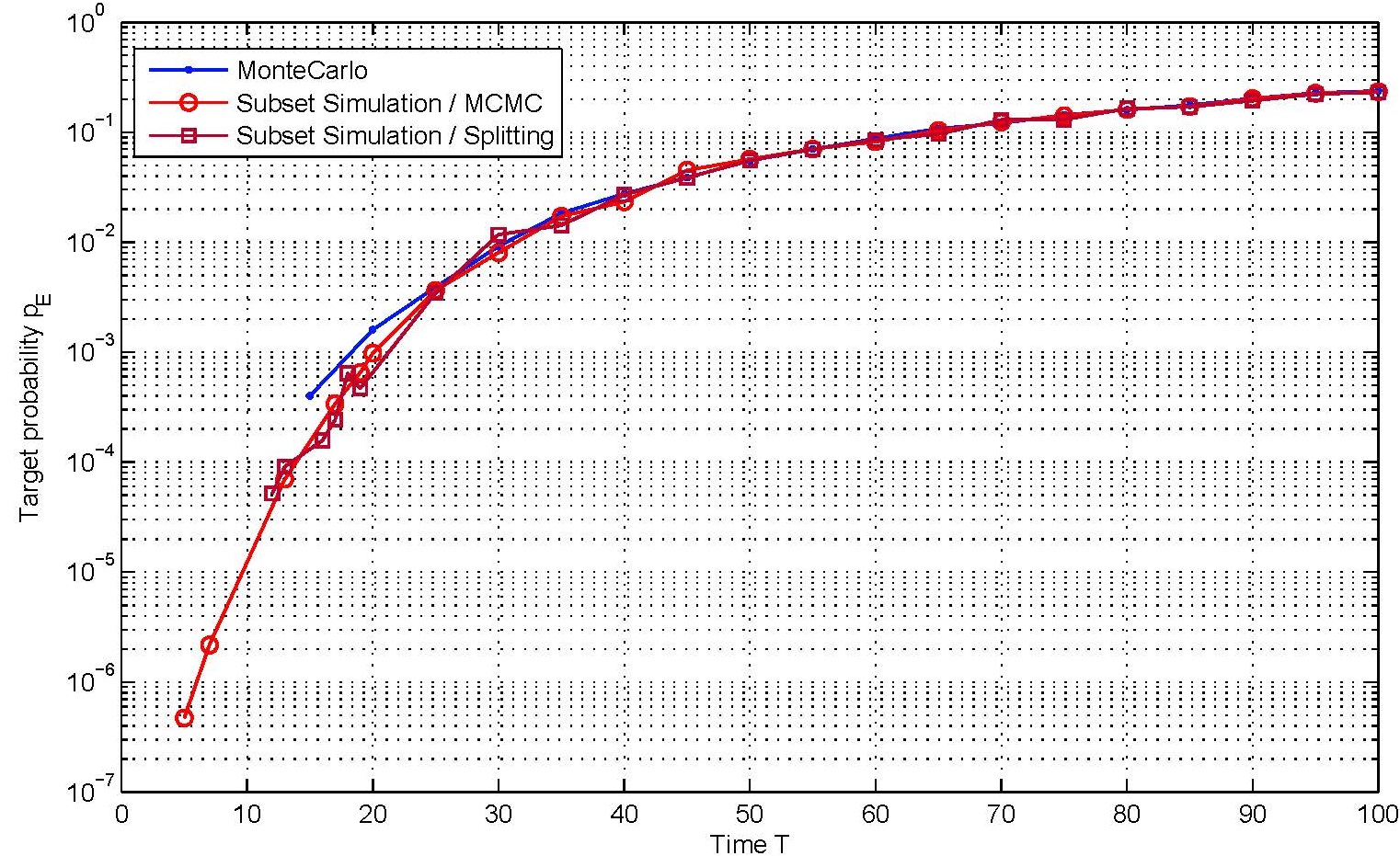}}
	\caption{\footnotesize Top panel shows the estimate of the probability $p_\mathcal{E}$ of event $\mathcal{E}$ where $\alpha=3$ as a function of duration time $T$. 
		For each value of $T\in[5,100]$, $N=10^4$ samples were used in MCS and $n=2\times10^3$ samples per conditional level were used in the two versions of SS. The MCS and SS/Splitting estimates for $p_\mathcal{E}$ are zero for $T<15$ and $T<12$, respectively. The bottom panel shows the total computational effort automatically chosen by both SS algorithms.} \label{fig2}
\end{figure}

\textit{Monte Carlo Simulation:} 
For $\alpha=3$, Figure~\ref{fig2} shows the probability $p_\mathcal{E}$ of event $\mathcal{E}$ as a function of $T$ estimated using standard MCS:  
\begin{equation}\label{eq:MCS_ex}
\widehat{p}_\mathcal{E}^{MCS}=\frac{1}{N}\sum_{i=1}^N I_{\mathcal{E}}(Z^{(i)}),
\end{equation}
where $Z^{(i)}=(Z_1^{(i)},\ldots,Z_D^{(i)})\sim\phi(z)$ are i.i.d. samples from the standard $D$-dimensional Gaussian PDF $\phi(z)$. For each value of $T$, $N=10^4$  samples were used. When $T<25$ the accuracy of the MCS estimate (\ref{eq:MCS_ex}) begins to degenerate since the total number of samples $N$ becomes too small for the corresponding target probability. Moreover, for $T<15$, none of the $N$ generated MCS samples belong to the target domain $\mathcal{E}$, making the MCS estimate zero. Figure~\ref{fig2} shows, as expected, that $p_\mathcal{E}$ is an increasing function of $T$, since the more time the system has, the more likely  its trajectory eventually penetrates the boundary of ellipsoid $\mathbb{E}$. 

\begin{figure}[t]
	\centerline{\includegraphics[angle=0,scale=0.45]{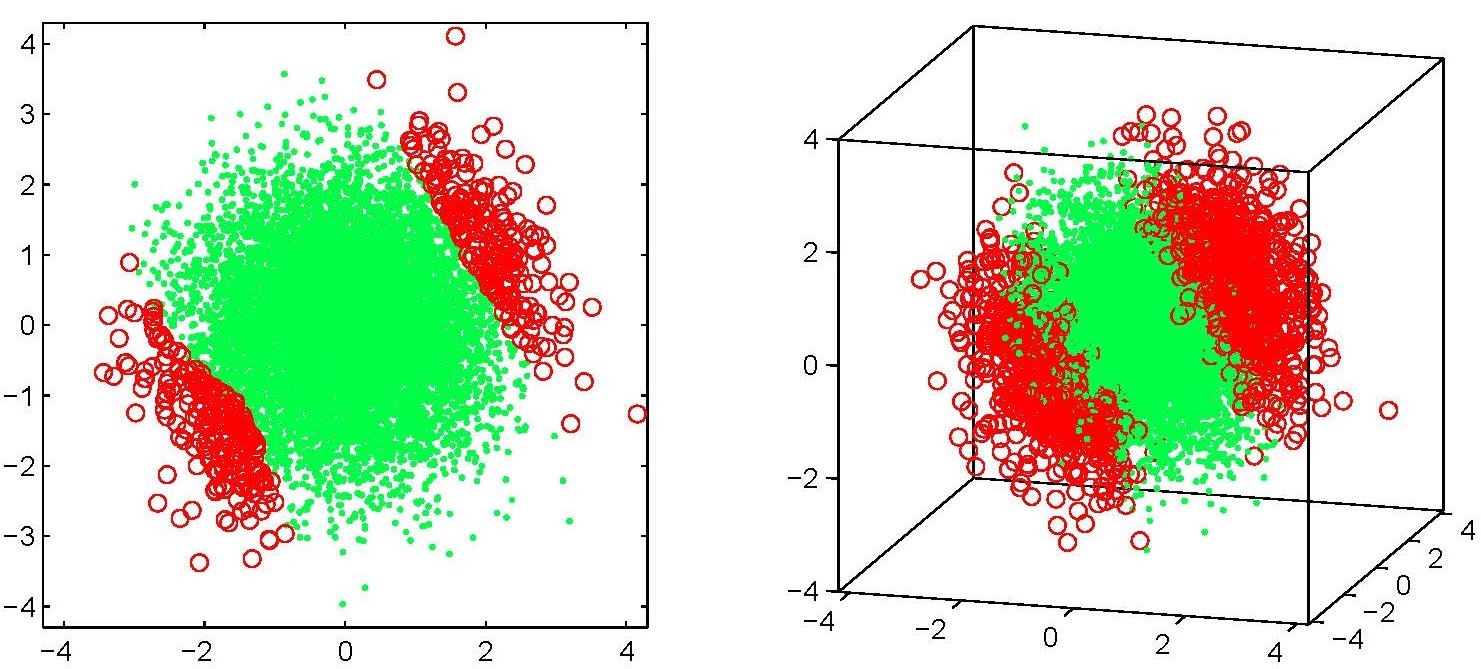}}
	\caption{\footnotesize Left panel: visualization of the  domain $\mathcal{E}$ in two dimensional case $D=2$, where $T=1$, $\Delta t=0.5$, and $\alpha=20$. $N=10^4$ samples were generated and marked by red circles (respectively, green dots) if they do (respectively, do not) belong to $\mathcal{E}$. Right panel: the same as on the left panel but with $D=3$ and $T=1.5$. } \label{fig3}
\end{figure}

\textit{Importance Sampling:}
IS is a variance reduction technique and, as it was discussed in previous sections, its efficiency critically depends on the choice of the ISD $q$. Usually some geometric information about the target domain $\mathcal{E}$ is needed for constructing a good ISD. To get some intuition, Figure~\ref{fig3} shows the domain $\mathcal{E}$ for two lower dimensional cases: $T=1$, $\Delta t=0.5$ ($D=2$) and $T=1.5$, $\Delta t=0.5$ ($D=3$). Notice that in both cases, $\mathcal{E}$ consists of two well separated subsets, $\mathcal{E}=\mathcal{E}_-\cup \mathcal{E}_+$, which are approximately symmetric about the origin. This suggests that a good ISD must be a mixture of two distributions $q_-$ and $q_+$, that effectively sample $\mathcal{E}_-$ and $\mathcal{E}_+$,
\begin{equation}
q(z)=\frac{q_-(z)+q_+(z)}{2}
\end{equation}
In this example, three different ISDs, denoted $q_1, q_2,$ and $q_3$, are considered: 
\begin{description}
	\item[Case 1:] $q_{\pm}(z)=\phi(z|\pm z_{\mathcal{E}})$, where $z_{\mathcal{E}}\sim\phi(z|\mathcal{E})$. That is, we first generate a sample $z_{\mathcal{E}}\in\mathcal{E}$ and then take ISD $q_1$ as the mixture of Gaussian PDFs centered at $z_{\mathcal{E}}$ and $-z_{\mathcal{E}}$. 
	\item[Case 2:] $q_{\pm}(z)=\phi(z|\pm z_{\mathcal{E}}^*)$, where $z_{\mathcal{E}}^*$ is obtained as follows. First we generate $n=1000$ samples from $\phi(z)$, and define $z_{\mathcal{E}}^*$ to be the sample in $\mathcal{E}$ with the smallest norm. Sample $z_{\mathcal{E}}^*$ can be interpreted as the ``best representative'' of $\mathcal{E}_-$ (or $\mathcal{E}_+$), since $\phi(z_{\mathcal{E}}^*)$ has the largest (among generated samples) value. We then take ISD $q_2$ as the mixture of Gaussian PDFs centered at $z_{\mathcal{E}}^*$ and $-z_{\mathcal{E}}^*$. 
	\item[Case 3:] To illustrate what happens if one ignores the geometric information about two components of $\mathcal{E}$, we choose $q_3(z)=\phi(z|z_{\mathcal{E}}^*)$, as given in Case 2.
\end{description}

Let $T=1$ and $\alpha=20$. The dimension of the uncertain excitation space is then $D=10$. Table~\ref{tab1} shows the simulation results for the above three cases as well as for standard MCS. The IS method with $q_1$, on average, correctly estimates $p_\mathcal{E}$. However the c.o.v. of the estimate is very large, which results in large fluctuations of the estimate in independent runs. IS with $q_2$ works very well and outperforms MCS: the c.o.v. is reduced by half. Finally, IS with $q_3$ completely misses one component part of the target domain $\mathcal{E}$, and the resulting estimate is about half of the correct value. Note that the c.o.v. in this case is very small, which is very misleading.  
\begin{table}[t]
	\caption{Simulation results for IS and MCS. For each method, mean values $\langle\widehat{p}_\mathcal{E}\rangle$ of the estimates and their coefficient of variations $\delta(\widehat{p}_\mathcal{E})$ are based on 100 independent runs.}
	\label{tab1}  
	\centering 
	\begin{tabular}{lll} 
		\hline\noalign{\smallskip}
		& $\langle\widehat{p}_\mathcal{E}\rangle$ & $\delta(\widehat{p}_\mathcal{E})$\\ 
		\noalign{\smallskip}\hline\noalign{\smallskip}
		MCS& $3.4\times10^{-3}$& 17$\%$  \\
		IS $q_1$ & $3.2\times10^{-3}$& 132.4$\%$ \\
		IS $q_2$ & $3.4\times10^{-3}$& 8.3$\%$  \\
		IS $q_3$ & $1.8\times10^{-3}$& 5.5$\%$ \\
	\noalign{\smallskip}\hline
	\end{tabular} 
\end{table}
\begin{figure}[t]
	\centerline{\includegraphics[angle=0,scale=0.4]{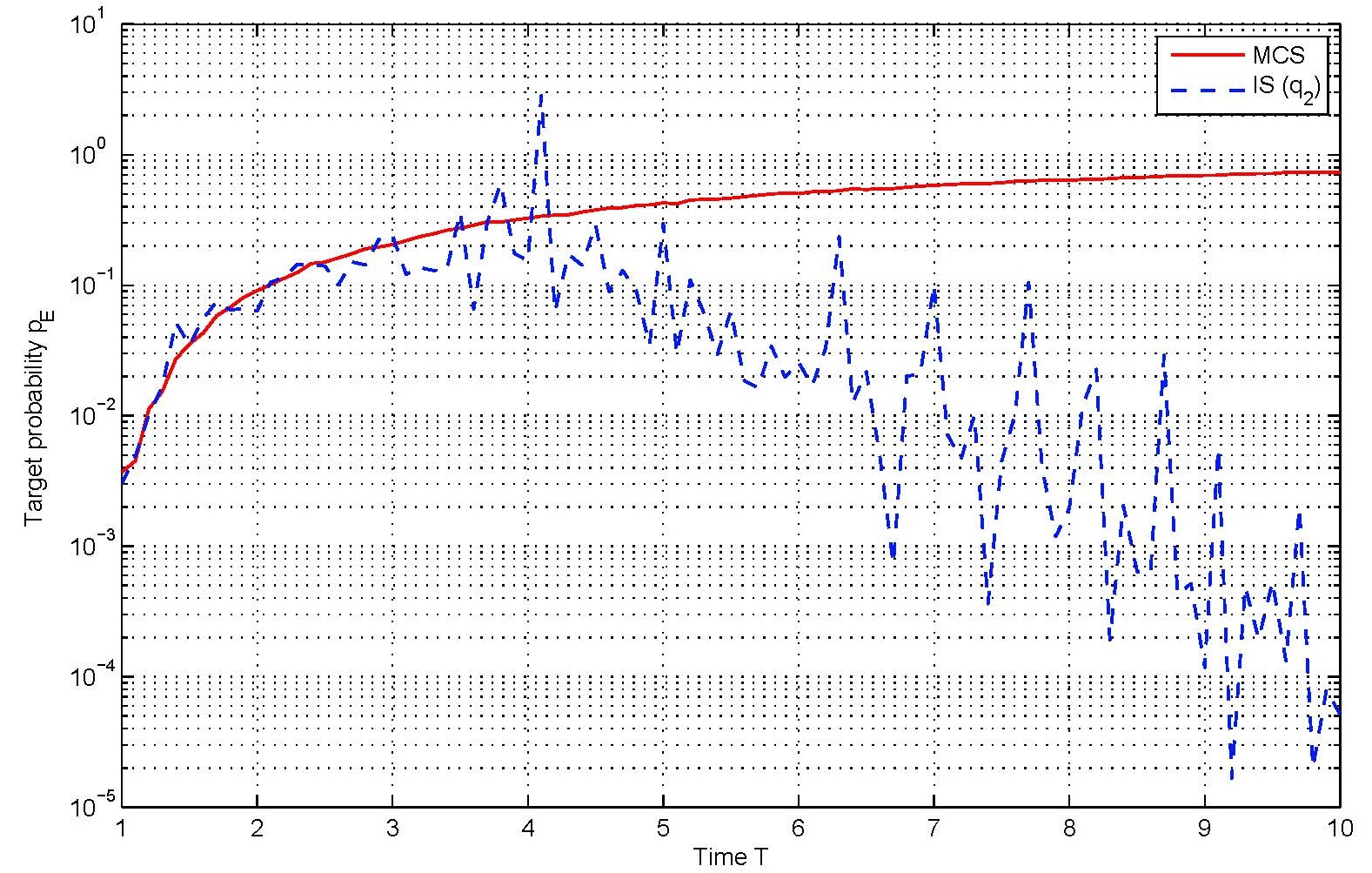}}
	\caption{\footnotesize Estimation of the target  probability $p_\mathcal{E}$ as a function of duration time $T$. Solid red and dashed blue curves correspond to MCS and IS with $q_2$, respectively. In this example,  $\alpha=20$ and $N=10^4$ samples for each value of $T$ is used. It is clearly visible how the IS estimate degenerates as the dimension $D$  goes from 10 ($T=1$) to 100 ($T=10$). } \label{fig4}
\end{figure}

It was mentioned in previous sections that IS is often not efficient in high dimensions because  it becomes more difficult to construct a good ISD \cite{IShighdim,KatZuev}. To illustrate this effect, IS with $q_2$ was used to estimate $p_\mathcal{E}$ for a sequence of problems where the total duration time gradually grows from $T=1$ to $T=10$. This results in an increase of the dimension $D$ of the underlying uncertain excitation space from 10 to 100. Figure~\ref{fig4} shows how the IS estimate degenerates as the dimension $D$ of the problem increases. While IS is accurate when $D=10$ ($T=1$), it strongly underestimates the true value of $p_\mathcal{E}$ as $D$ approaches $100$ ($T=10$).

\textit{Subset Simulation:}
SS is a more advanced simulation method and, unlike IS, it does not suffer from the curse of dimensionality. For $\alpha=3$, Figure~\ref{fig2} shows the estimate of the target probability $p_\mathcal{E}$ as a function of $T$  using  SS with MCMC and Splitting. For each value of $T$, $n=2\times10^3$ samples were used in each conditional level in SS.  Unlike MCS, SS is capable of efficiently simulating very rare events and estimating their small probabilities. The total computational effort, i.e. the total number $N$ of samples automatically chosen by SS, is shown in the bottom panel of Figure~\ref{fig2}. Note that the larger the value of $p_\mathcal{E}$, the smaller the number of conditional levels in SS, and, therefore, the smaller the total number of samples $N$. The total computational effort in SS is thus a decreasing function of $T$. In this example, the original MCMC strategy \cite{SS} for generating conditional samples outperforms the splitting strategy \cite{ChingAuBeck} that exploits the causality of the system: while the SS/MCMC method works even in the most extreme case  ($T=5$), the SS/Splitting estimate for $p_\mathcal{E}$ becomes zero for $T<12$.

\section{Conclusion}
This chapter examines computational methods for rare-event simulation in the context of uncertainty quantification for dynamic systems that are subject to future uncertain excitation modeled as a stochastic process. The rare events are assumed to correspond to some time-varying performance quantity exceeding a specified threshold over a specified time duration, which usually means that the system performance fails to meet some design or operation specifications. 

To analyze the reliability of the system against this performance failure, a computational model for the input-output behavior of the system is used to predict the performance of interest as a function of the input stochastic process discretized in time. This dynamic model may involve explicit treatment of parametric and non-parametric uncertainties that arise because the model only approximately describes the real system’s behavior, implying that there are usually no true values of the model parameters and the accuracy of its predictions are uncertain. In the engineering literature, the mathematical problem to be solved numerically for the probability of performance failure, commonly called the failure probability, is referred to as the first-passage reliability problem. It does not have an analytical solution and numerical solutions must face two challenging aspects:
\begin{enumerate}
	\item The vector representing the time-discretized stochastic process that models the future system excitation lies in an input space of high dimension;
	\item The dynamic systems of interest are assume to be highly reliable so that their performance failure is a rare event, that is, the probability of its occurrence, $p_\mathcal{E}$, is very small.
\end{enumerate}
As a result, standard Monte Carlo Simulation and Importance Sampling methods are not computationally efficient for first-passage reliability problems. On the other hand, Subset Simulation has proved to be a general and powerful method for numerical solution of these problems. Like MCS, it is not affected by the dimension of the input space and for a single run, it produces a plot of $p_\mathcal{E}$ vs threshold $b$ covering $p_\mathcal{E}\in[p_0^{-L},1]$, where $L$ is the number of levels used. For a critical appraisal of methods for first-passage reliability problems in high dimensions, the reader may wish to check Schu\"{e}ller et al \cite{SPK04}.

Several variants of Subset Simulation have been developed motivated by the goal of further improving the computational efficiency of the original version, although the efficiency gains, if any, are modest. All of them have an accuracy described by a coefficient of variation for the estimate of the rare-event probability that depends on $\ln(1/p_\mathcal{E})$ rather than $\sqrt{1/p_\mathcal{E}}$ as in standard Monte Carlo simulation. For all methods covered in this section, the dependence of this coefficient of variation on the number of samples $N$ is proportional to $N^{-1/2}$. Therefore, in the case of very low probabilities, $p_\mathcal{E}$, it still requires thousands of simulations (large $N$) of the response time history based on a dynamic model as in (\ref{model}) in order to get acceptable accuracy. For complex models, this computational effort may be prohibitive. 

One approach to reduce the computational effort when estimating very low rare-event probabilities is to utilize additional information about the nature of the problem for specific classes of reliability problems (e.g. \cite{Au,AuBeck2001}). Another more general approach is to construct surrogate models (meta-models) based on using a relatively small number of complex-model simulations as training data. The idea is to use a trained surrogate model to rapidly calculate an approximation of the response of the complex computational model as a substitute when drawing new samples. Various methods for constructing surrogate models have been applied in reliability engineering, including response surfaces \cite{BB90}, support vector machines \cite{BDL11,Hu04},  neural networks \cite{PGLP12}, and Gaussian process modeling (Kriging) \cite{DSD13}. The latter method is a particularly powerful one because it also provides a probabilistic assessment of the approximation error. It deserves further exploration, especially with regard to the optimal balance between the accuracy of the surrogate model as a function of the number of training samples from the complex model, and the accuracy of the estimate of the rare-event probability as a function of the total number of samples from both the complex model and the surrogate model.


\end{document}